\documentclass[aps,prc,twocolumn,preprintnumbers,superscriptaddress,showpacs,showkeys]{revtex4}

\usepackage{graphicx}
\usepackage{amsmath}

\newcommand{\sqrtsNN}{\mbox{$\sqrt{\mathrm{s}_{_{\mathrm{NN}}}}$}}
\newcommand{\alam}{\overline{\Lambda}}
\newcommand{\lam}{\Lambda}
\newcommand{\kp}{K^+}
\newcommand{\km}{K^-}
\newcommand{\axi}{\overline{\Xi}^+}
\newcommand{\xim}{\Xi^-}
\newcommand{\xiz}{\Xi^0}
\newcommand{\omm}{\Omega^-}
\newcommand{\aom}{\overline{\Omega}^+}
\newcommand{\lratio}{${\alam}/{\lam}$}
\newcommand{\xratio}{${\axi}/{\xim}$}
\newcommand{\kratio}{${\kp}/{\km}$}
\newcommand{\pratio}{${\overline{p}}/{p}$}
\newcommand{\omratio}{${\aom}/{\omm}$}
\newcommand{\dedx}{$dE/dx$}
\newcommand{\pt}{$p_T$}

\newcommand{\xival}{0.83}
\newcommand{\xierr}{0.04}
\newcommand{\xisys}{0.05}
\newcommand{\lamval}{0.71}
\newcommand{\lamerr}{0.01}
\newcommand{\lamsys}{0.04}
\newcommand{\kval}{1.075}
\newcommand{\kerr}{0.008}
\newcommand{\ksys}{0.03}
\newcommand{\kinkval}{1.13}
\newcommand{\kinkerr}{0.015}
\newcommand{\kinksys}{0.05}
\newcommand{\kcombval}{1.092}
\newcommand{\kcomberr}{0.023}
\newcommand{\omval}{0.95}
\newcommand{\omerr}{0.15}
\newcommand{\omsys}{0.05}

\begin{document}


\title{Strange anti-particle to particle ratios at mid-rapidity in\\
\sqrtsNN\ =~130~GeV Au+Au Collisions}

%

\affiliation{Argonne National Laboratory, Argonne, Illinois 60439}

\affiliation{Brookhaven National Laboratory, Upton,New York 11973}

\affiliation{University of Birmingham, Birmingham, United Kingdom}

\affiliation{University of California, Berkeley, California 94720}

\affiliation{University of California, Davis, California 95616}

\affiliation{University of California, Los Angeles, California 90095}

\affiliation{Carnegie Mellon University, Pittsburgh, Pennsylvania 15213}

\affiliation{Creighton University, Omaha, Nebraska 68178}

\affiliation{Laboratory for High Energy (JINR), Dubna, Russia}

\affiliation{Particle Physics Laboratory (JINR), Dubna, Russia}

\affiliation{University of Frankfurt, Frankfurt, Germany}

\affiliation{Indiana University, Bloomington, Indiana 47408}

\affiliation{Institut de Recherches Subatomiques, Strasbourg, France}

\affiliation{Kent State University, Kent, Ohio 44242}

\affiliation{Lawrence Berkeley National Laboratory, Berkeley, California 94720}

\affiliation{Max-Planck-Institut fuer Physik, Munich, Germany}

\affiliation{Michigan State University, East Lansing, Michigan 48824}

\affiliation{Moscow Engineering Physics Institute, Moscow Russia}

\affiliation{City College of New York, New York City, New York 10031}

\affiliation{Ohio State University, Columbus, Ohio 43210}

\affiliation{Pennsylvania State University, University Park, Pennsylvania 16802}

\affiliation{Institute of High Energy Physics, Protvino, Russia}

\affiliation{Purdue University, West Lafayette, Indiana 47907}

\affiliation{Rice University, Houston, Texas 77251}

\affiliation{Universidade de Sao Paulo, Sao Paulo, Brazil}

\affiliation{SUBATECH, Nantes, France}

\affiliation{Texas A \& M, College Station, Texas 77843}

\affiliation{University of Texas, Austin, Texas 78712}

\affiliation{Warsaw University of Technology, Warsaw, Poland}

\affiliation{University of Washington, Seattle, Washington 98195}

\affiliation{Wayne State University, Detroit, Michigan 48201}

\affiliation{Institute of Particle Physics, CCNU (HZNU), Wuhan, 430079 China}

\affiliation{Yale University, New Haven, Connecticut 06520}

\author{J.~Adams}\affiliation{University of Birmingham, Birmingham, United Kingdom}
\author{C.~Adler}\affiliation{University of Frankfurt, Frankfurt, Germany}
\author{Z.~Ahammed}\affiliation{Purdue University, West Lafayette, Indiana 47907}
\author{C.~Allgower}\affiliation{Indiana University, Bloomington, Indiana 47408}
\author{J.~Amonett}\affiliation{Kent State University, Kent, Ohio 44242}
\author{B.D.~Anderson}\affiliation{Kent State University, Kent, Ohio 44242}
\author{M.~Anderson}\affiliation{University of California, Davis, California 95616}
\author{G.S.~Averichev}\affiliation{Laboratory for High Energy (JINR), Dubna, Russia}
\author{J.~Balewski}\affiliation{Indiana University, Bloomington, Indiana 47408}
\author{O.~Barannikova}\affiliation{Purdue University, West Lafayette, Indiana 47907}\affiliation{Laboratory for High Energy (JINR), Dubna, Russia}
% ...

\collaboration{STAR Collaboration}\homepage{www.star.bnl.gov}\noaffiliation

\date{\today}

\begin{abstract}
Values of the ratios in the mid-rapidity yields of \lratio\
 = $\lamval\pm\lamerr(stat.)\pm\lamsys(sys.)$,\ \xratio = $\xival\pm\xierr(stat.)\pm\xisys(sys.)$, \omratio = $\omval\pm\omerr(stat.)\pm\omsys(sys.)$ and \kratio =
 $\kcombval\pm\kcomberr(combined)$ were obtained in central \sqrtsNN\ =~130~GeV Au+Au collisions using the STAR detector. The ratios
 indicate that a fraction of the net-baryon number from the initial system is present in the excess of hyperons
 over anti-hyperons at mid-rapidity.
 The trend in the progression of the baryon ratios, with increasing
 strange quark content, is similar to that observed in heavy-ion collisions at lower energies. The value of these ratios
 may be related to the charged kaon ratio in the framework of simple quark-counting and thermal models.
\end{abstract}


\maketitle

The goal of the experimental program at the Relativistic Heavy Ion Collider (RHIC) is to study new states of
nuclear matter which have been predicted to form in heavy-ion collisions~\cite{Blaizot:1999bv,Specht:2002qx} and
for which much indirect evidence has emerged from a previous series of experiments at lower energy
\cite{Beker:1995qv,Aggarwal:2000th,Albrecht:1996fs,Andersen:1999ym,Abreu:1997jh,Abreu:2000ni,Appelshauser:1998rr,Margetis:1995tt,
Bearden:1997dd}. Measurements of anti-particle to particle ratios in these collisions give information on the net
baryon density or baryochemical potential achieved~\cite{Braun-Munzinger:1999qy} and are thus of interest in
characterizing the environment created in these collisions. It has also been suggested that the measurement of
strange anti-baryon to baryon ratios could help distinguish between a hadron gas and a deconfined plasma of quarks
and gluons~\cite{Koch:1988db}. The dominant production mechanism for anti-quarks is via gluon
fusion~\cite{Koch:1986ud,Rafelski:1982pu} and a measurement of the anti-baryon to baryon ratio therefore probes
the gluonic degrees of freedom. The relations between the various anti-particle to particle ratios allow for the
test of a non-linear quark coalescence model~\cite{Bialas:1998ea,Zimanyi:2000ky} which is consistent with the
existence of quark degrees of freedom. We present here the first measurements of multi-strange baryon production
at \sqrtsNN\ =~130~GeV and utilize recently revised \pratio\ \cite{Adler:2003Erratum} and published \lratio\
\cite{Adler:2002pba} results to compare to models.

The Solenoidal Tracker at RHIC (STAR) detector system~\cite{Ackermann:1999kc}, in the configuration used to
collect the data presented
 here, consisted principally of a large cylindrical Time Projection Chamber (TPC) used for charged particle tracking.
 The TPC has inner and outer radii of 50~cm and 200~cm respectively, a total length of approximately 420~cm and was operated in a
 0.25~Tesla magnetic field. It is surrounded by a cylinder of scintillator slats forming a Central Trigger Barrel
 (CTB),  a fast detector providing a signal proportional to the multiplicity within pseudo-rapidity $\pm1$. Two Zero Degree Calorimeters (ZDCs) were used to detect
 spectator neutrons from the colliding ions at close to beam rapidities \cite{Adler:2000bd}. Collisions were triggered by requiring coincident
 signals in the ZDCs which formed a minimum bias trigger. Approximately 250,000 of these events were used in the
 analysis.
An enriched central data sample was acquired, with the additional requirement of a high CTB threshold,
corresponding approximately to the $14\%$ most central events. On these events a further centrality selection was
made off-line, by cutting on the observed track multiplicity in the TPC after event reconstruction. This analysis
used approximately 180,000 central \sqrtsNN\ =~130~GeV Au+Au events after the multiplicity cut, corresponding to
the most central $11\%$ of the total hadronic cross-section~\cite{Ackermann:2000tr}.

Two techniques were used to extract the raw yields of strange particles. First, charged kaons were identified via
their specific ionization, or energy loss (\dedx), in the TPC. Second, these and other strange particles were
reconstructed from their decay topology.

Up to 45 ionization samples were measured for each track. The \dedx\ resolution was measured to be 11$\%$
following the procedure of~\cite{Aguilar-Benitez:1991yy}. For the charged kaon \dedx\ analysis only tracks below a
momentum of 0.6~GeV/$c$ were used, where the \dedx\ of kaons is distinct from those of other particle species. In
addition, tracks were required to originate from the primary interaction vertex within 3 cm. Similar
to~\cite{Adler:2001bp}, the distribution of $Z=log[(dE/dx)_{Meas}/(dE/dx)_{BB}]$, where $(dE/dx)_{BB}$ is the
\dedx\ from a Bethe-Bloch parameterization, is fitted with a convolution of Gaussian functions. The kaon raw
yields were extracted from the fit results for each \pt\ bin within rapidity $|y| < 0.4$.

\begin{figure}
\includegraphics[width=0.5\textwidth]{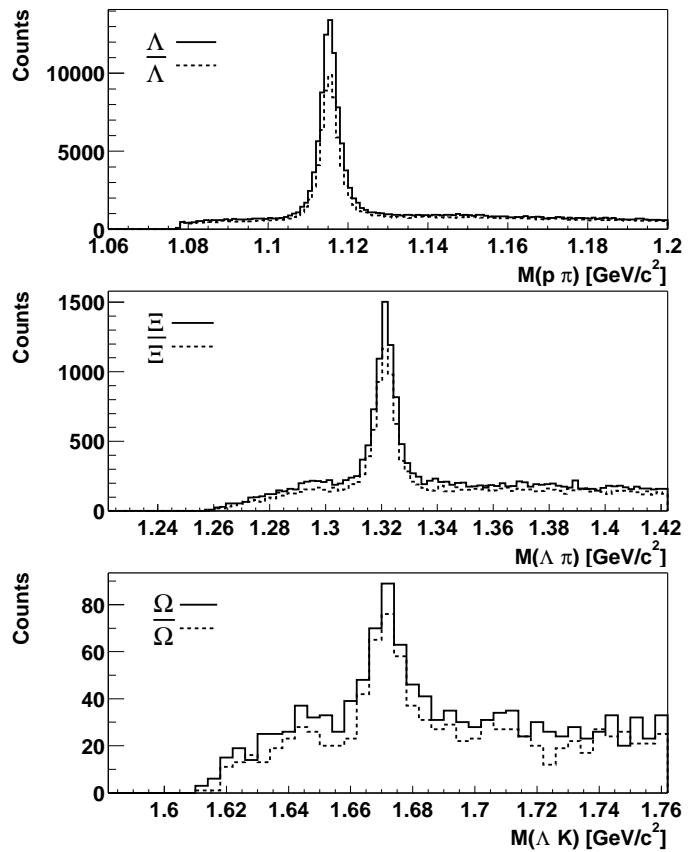}
\caption{Invariant mass distributions for $p\pi^-$ and $\overline{p}\pi^+$ (top panel), $\lam\pi^-$ and
$\alam\pi^+$ (middle panel) and $\lam K^-$ and $\alam K^+$ (lower panel).\label{fMasses}}
\end{figure}

The most versatile technique for the reconstruction of strange particles is via their decay
topologies~\cite{Margetis:2000sv}. The decay $\lam \rightarrow p \pi^-$ (64$\%$ branching ratio) and the charge
conjugate decay for $\alam$ result in two charged particles in the final state. The $\lam$ particles from the
electro-magnetic decay of the $\Sigma^{0}$ are included in the $\lam$ sample since they were not experimentally
distinguishable from the primary $\lam$ population. The momenta of these charged daughter particles are calculated
from their trajectories in the TPC. Both tracks can be extrapolated back toward the primary interaction vertex to
locate their common point of origin, where the kinematic properties of the parent can be calculated. In this
process, all pairs of positively and negatively charged tracks in an event are considered. To reduce the large
combinatorial background which results from random crossings of tracks interior to the TPC fiducial volume,
additional cuts must be made. The most important criteria for improving the signal to noise ratio are that the
 decay vertex is well separated from the primary interaction and that the parent originates from the primary interaction
 vertex while the daughters do not. An additional requirement that the \dedx\ of the daughters is compatible with the
 expected decay mode is also applied. For example the positively charged daughter of a $\lam$ should have a \dedx\
 compatible with it being a proton. This helps to suppress the background from fake decays with two pion daughter particles with $p\lesssim$ 1 GeV/c.
  This technique is extended to enable the reconstruction of the $\xim \rightarrow \lam \pi^-$
  and $\omm \rightarrow \lam \km$ decays ($100\%$ and
   $68\%$ branching ratios respectively) and their charge conjugate $\axi$ and $\aom$ decays. Here, only $\lam$ (or
$\alam$) candidates within $\pm7$~MeV/$c^2$ of the expected mass~\cite{Hagiwara:2002pw} are used and the
 requirement that the $\lam$ (or $\alam$) originates from the primary interaction vertex is relaxed. The resulting
  invariant mass distributions for $\lam$, $\xim$ and $\omm$, with their anti-particle distributions superimposed, are shown in
  Figure \ref{fMasses}. The remaining background under the peak in each invariant mass
  distribution was subtracted by using a linear interpolation between the background regions a few MeV/$c^2$ on either side
  of the peak region.

Charged kaons can also be reconstructed using a variation on this topological technique via their one-prong decay
channels. The most prominent of these are $K \rightarrow \mu \nu$\ and $K \rightarrow \pi \pi^0$, with $64\%$ and
$21\%$ branching ratios, respectively. In this ``kink'' method the tracks from the charged kaon and charged
daughter particle are used to reconstruct the kinematics of the decay. In order that both parent and daughter are
reconstructed in the TPC with good momentum resolution, the fiducial volume for the location of the decay vertex
is restricted to radii of $130-180$ cm. The background comes from charged pion decays, multiple scattering and
hadronic interactions in the TPC gas and combinatorics. The pion decay contribution can be largely eliminated by a
cut on the opening angle between the parent and daughter tracks. This angle, for a given momentum, is much smaller
for a pion decay than a kaon decay. The remaining background level was estimated to be approximately
15$\%$~\cite{DengThesis}. The method allows charged kaons to be identified over a wide range in \pt.

The central assumption in forming the anti-particle to particle ratios is that the detector response is symmetric
with respect to charge and therefore no corrections to the yields for the reconstruction efficiency or detector
acceptance are required. However, losses due to the absorption of anti-protons in the detector material have the
potential to modify the \lratio, \xratio\ and \omratio\ ratios and feed-down from the decay of heavier strange
baryons can modify both the \lratio\ and \xratio\ ratios.
 Absorption causes the final state anti-proton from $\alam$ decay
to fail to be reconstructed more often than the proton from $\lam$ decay. The size of this effect has been
estimated and corrected for using a GEANT simulation of the detector. Absorption reduces the apparent \lratio\ and
\xratio\ ratios by $1\%$ and $0.2\%$ respectively. The decaying anti-particles also have a larger absorption
cross-section than their corresponding particles, but since they decay within a few centimeters, before most of
the absorbing materials have been traversed, this correction is even smaller and is implicitly included in the
numbers given above. The observed \lratio\ includes feed-down contributions.  The total $\lam$ yield contains
$\lam$ originating from $\xim$, $\xiz$ and $\omm$ decays, estimated to be $27\pm6\%$~\cite{Adler:2002pba}. Their
anti-particle decays similarly contribute to $\alam$. Assuming that these feed-down contributions to $\lam$ and
$\alam$ are in the ratio of the \xratio\ measurement, we obtain an actual \lratio\ ratio, which we quote,
$0.05-0.015$ lower than the observed value. The only feed-down contribution to the \xratio\ comes from the $\omm
\rightarrow \xim \pi^0$ channel with a $9\%$ branching ratio and was therefore neglected. Two processes modify the
\kratio\ ratio. Feed-down of kaons from the decay $\phi \rightarrow K^+K^-$ was estimated, using the measured
$\phi/K$ ratio \cite{Adler:2002xv,Adler:2002wn}, to reduce the apparent \kratio\ ratio by $0.8\%$ at \pt\ = 0.4
GeV/$c$ and less than $0.3\%$ above \pt\ = 1 GeV/$c$. Secondary interactions were studied using GEANT simulations
of HIJING~\cite{Wang:1997yf} events and were found to increase the measured ratio by $0.7\%$. These corrections
were applied in producing the final ratio.

\begin{figure}
\includegraphics[width=0.5\textwidth]{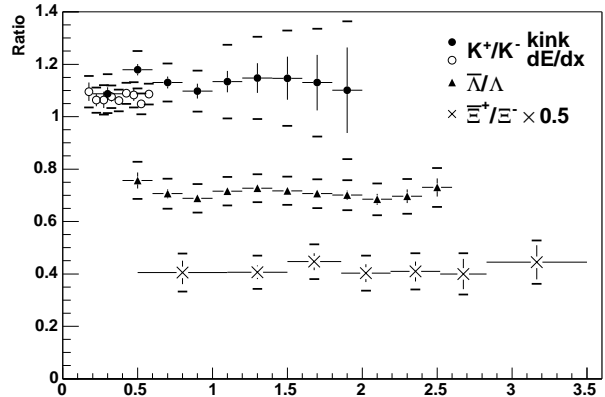}
\caption{The ratios \lratio, \kratio\ and \xratio\ as a function of \pt. The error bars indicate the statistical
errors and the brackets the systematic uncertainties. 
\label{fPt}}
\end{figure}

 After the absorption correction the value of the \lratio\ ratio is $0.74\pm0.01(stat.)$ in the measured
 acceptance interval of
$p_T>0.4$ GeV/$c$ and within one unit of rapidity centered at mid-rapidity. This is the same is the value for the
corrected data reported previously \cite{Adler:2002pba}. The feed-down correction reduces this to a final value of
$\lamval\pm\lamerr(stat.)$. The \xratio\ ratio after correction is $\xival\pm\xierr(stat.)$, measured over the
same rapidity interval and $p_T>0.5$ GeV/$c$. In order to admit a larger sample of $\omm$ and $\aom$, the
\omratio\ ratio was calculated using a larger interval, of $\pm1$ units of rapidity, and a value of
$\omval\pm\omerr(stat.)$ was obtained. The \lratio\ and \xratio\ ratios as a function of \pt\ out to 2.5 GeV/$c$
and 3.5 GeV/$c$ are shown in Figure \ref{fPt} and are consistent with a constant value. Within statistics the
\lratio\ ratio appears to be independent of the charged particle yield at mid-rapidity. Systematic uncertainties
on the \lratio\ and \xratio\ of \lamsys\ and \xisys\ respectively have been estimated by varying the cuts used to
identify decay candidates. There were insufficient data to estimate the systematic uncertainty on the \omratio\
this way so the \xratio\ systematic uncertainty was used since the reconstruction methods are identical. The
\kratio\ ratio is $\kval\pm\kerr(stat.)$, measured via the \dedx\ method, in the range $0.15<p_T<0.6$ GeV/$c$ and
$\pm0.4$ units of rapidity around mid-rapidity. The same ratio measured by the kink method is
$\kinkval\pm\kinkerr(stat.)$ and extends out to 2 GeV/$c$ in \pt. The systematic error of the kink measurement due
to detector effects is estimated to be $\kinksys$ and that for the \dedx\ is estimated at $\ksys$. The small
discrepancy in the \kratio\ from the two methods is within the estimated systematic errors and a combined value of
$\kcombval\pm\kcomberr$ was calculated, following the method of the PDG~\cite{Hagiwara:2002pw} when combining
results from different experiments. As Figure \ref{fPt} shows, with both methods, the ratio shows no significant
deviation from a constant as a function of \pt. As all the \pt\ intervals cover a large fraction of the total
yield (over $70\%$ \cite{Adler:2002pba,Castillo:2002xv,Adler:2002wn}) we assume that the ratios we measure are a
good indication of the ratios in the integrated yields.

\begin{figure}
\includegraphics[width=0.5\textwidth]{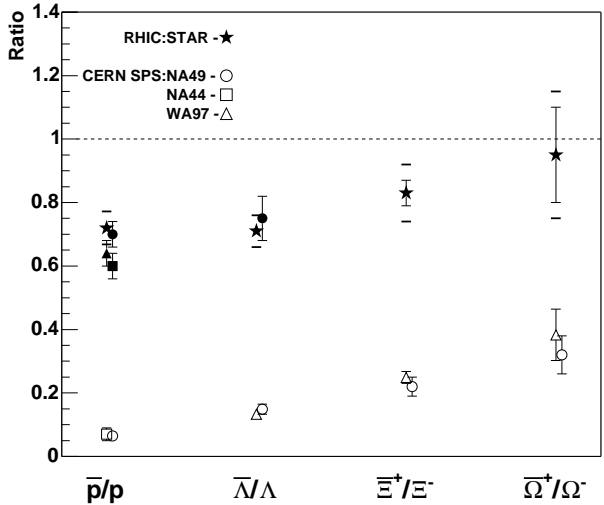}
\caption{Anti-baryon to baryon ratios measured by STAR and other RHIC experiments
\cite{Adcox:2001mf,Adcox:2002au,Bearden:2001kt,Back:2001qr}, for baryons of increasing strangeness content,
compared to values obtained in
experiments~\cite{Andersen:1999gd,Kaneta:1997qf,Afanasiev:2002fk,Afanasiev:2002he,Afanasiev:2002mx,Afanasiev:2002ub}
at the SPS. For STAR points the bars indicate the statistical uncertainties and the brackets the additional
systematic errors. \label{fRatiobaryons}}
\end{figure}

The strange anti-baryon to baryon ratios are plotted in Figure~\ref{fRatiobaryons} together with their values
found in central Pb+Pb collisions at \sqrtsNN\ =~17 GeV~\cite{Andersen:1999gd} at the CERN Super Proton
Synchrotron. Also shown are the \pratio\ ratios from STAR~\cite{Adler:2003Erratum} and two measurements at the
lower energy from NA44~\cite{Kaneta:1997qf} and NA49~\cite{Afanasiev:2002fk}. Figure~\ref{fRatiobaryons} shows
that the ratios increase with increasing strangeness content of the baryon at both \sqrtsNN\ =~17~GeV and
\sqrtsNN\ =~130~GeV. The increasing trend in the ratios may be explained in a simple quark coalescence
model~\cite{Bialas:1998ea,Zimanyi:2000ky}, which predicts that the anti-baryon to baryon ratios should be related
to one another by a common multiplicative factor. The multiplicative factor is given by the value of the $\kp/\km$
ratio. This is in approximate agreement with the data presented here, as shown in Table~\ref{tDouble}.

\begin{table}
\caption{\label{tDouble}\kratio\ ratio compared to compound ratios having the same net quark content. Comparisons
made for this experiment and experiments at
SPS~\cite{Kaneta:1997qf,Afanasiev:2002fk,Afanasiev:2002he,Afanasiev:2002mx,Afanasiev:2002ub,Andersen:1999gd}.}
\begin{ruledtabular}
\begin{tabular}{ccc}
 & STAR & SPS \\
\\
\kratio\ & $\kcombval\pm\kcomberr$ & $1.76\pm0.06$ \\
\\
$\dfrac{\alam / \lam}{\overline{p}/p}$ & $0.98\pm0.09$& $2.07\pm0.21$ \\
\\
$\dfrac{\axi / \xim} {\alam / \lam} $ & $1.17\pm0.11$& $1.78\pm0.15$ \\
\\
$\dfrac{\aom / \omm} {\axi / \xim} $ & $1.14\pm0.21$ & $1.42\pm0.22$ \\
\end{tabular}
\end{ruledtabular}
\end{table}

 Within the coalescence
model hadrons are formed from a system of independent quarks and anti-quarks. An alternative description of
particle production which nevertheless gives equivalent predictions for the ratios discussed here is the
statistical model approach~\cite{Letessier:1992rs}, which does not distinguish between quark or hadron degrees of
freedom. In this case the multiplicative factor is exp$(2\mu_B/3T - 2\mu_s/T)$, where $\mu_B$ is the baryon
chemical potential, $\mu_s$ is the strange quark chemical potential and $T$ is the chemical freeze-out
temperature. The $\kp/\km$ ratio can therefore be interpreted as an indirect measure of the baryon chemical
potential. If the central region in Au+Au collisions at \sqrtsNN\ =~130~GeV were net baryon free ($\mu_B = 0$),
then the $\kp/\km$ ratio would be equal to one, and in both models the anti-baryon to baryon ratios would also be
equal to one, under the assumption that strangeness is locally conserved. However, while the anti-baryon to baryon
ratios at this higher energy are closer to unity, reflecting a lower net-baryon density, this density is
nevertheless still positive. This is thought to be a consequence of baryon number transport (or stopping) during
the collision process. There is an excess of $u$ and $d$ quarks over their anti-quarks favoring the production of
baryons over anti-baryons and $\kp$ over $\km$. We find, from a fit to all the ratios, that $\mu_B/T =
0.18\pm0.03$ and $\mu_s/T = 0.001\pm0.011$ with $\chi^{2}/dof = 2.5$ when including all the systematic errors. A
statistical model analysis using preliminary data~\cite{Braun-Munzinger:2001ip} is also consistent giving $\mu_B/T
= 0.26\pm0.03$ where $\mu_B = 45$~MeV and $T=170$~MeV. This compares to $\mu_B/T = 1.58\pm0.04$ at \sqrtsNN\
=~17~GeV where $\mu_B = 266$~MeV and $T=168$~MeV~\cite{Braun-Munzinger:1999qy}. We also note that the flatness of
the \lratio\ and \kratio\ ratios in Figure \ref{fPt} suggests that the transverse momentum distributions of the
particles and their anti-particles are very similar. The matching \pt\ distributions are especially interesting
for the $\lam$ and $\alam$, since there may be different production mechanisms. The $\lam$ are believed to have
component due to associated production (e.g. $pp \rightarrow p \lam K$) by the incoming baryons.

 In summary, we have reported
strange anti-particle to particle ratios measured by the STAR experiment at mid-rapidity in the $11\%$ most
central Au+Au collisions at \sqrtsNN\ =~130~GeV. The ratios
 indicate that a fraction of the net-baryon number from the initial system is present in the excess of hyperons
 over anti-hyperons at mid-rapidity. The ratios are consistent with
simple quark counting models and with a statistical description of particle production which is governed by a
common baryon chemical potential and chemical freeze-out temperature.



%



\begin{acknowledgments}
We wish to thank the RHIC Operations Group and the RHIC Computing Facility at Brookhaven National Laboratory, and
the National Energy Research Scientific Computing Center at Lawrence Berkeley National Laboratory for their
support. This work was supported by the Division of Nuclear Physics and the Division of High Energy Physics of the
Office of Science of the U.S. Department of Energy, the United States National Science Foundation, the
Bundesministerium fuer Bildung und Forschung of Germany, the Institut National de la Physique Nucleaire et de la
Physique des Particules of France, the United Kingdom Engineering and Physical Sciences Research Council, Fundacao
de Amparo a Pesquisa do Estado de Sao Paulo, Brazil, the Russian Ministry of Science and Technology, the Ministry
of Education of China, the National Natural Science Foundation of China, and the Swiss National Science
Foundation.
\end{acknowledgments}

\bibliography{strangeratio}

\end{document}